\documentclass[arxiv,usenatbib]{iupartex}

\usepackage{newtxtext,newtxmath}
\usepackage{framed,multirow}
\usepackage[T1]{fontenc}
\usepackage{mhchem}
\usepackage{float} 
\usepackage{amssymb}
\usepackage{latexsym}
\usepackage{amsmath}
\usepackage{graphicx}  
\usepackage{lscape}    
\usepackage{url}

\title[HabSim: Disruption Modeling Architecture]{HabSim: Architecture for Modelling Disruptions, Propagation, Detection and Repair in Deep Space Habitats}

\author[Vaccino et al.]{%
Luca Vaccino$^{1\cc}$
Alana K. Lund$^{2}$
Shirley J. Dyke$^{1}$
Mohsen Azimi$^{3}$
and
Ethan Vallerga$^{4}$
\affsep \\
$^1$Purdue University, School of Mechanical Engineering, West Lafayette, IN 47907, USA\\
$^2$University of Waterloo, Civil and Environmental Engineering Department, Waterloo, ON N2L 3G1, Canada\\
$^3$Department of Mechanical Engineering, Mississippi State University, Mississippi State, MS 39762, USA\\
$^4$Purdue University, School of Aeronautics and Astronautics, West Lafayette, IN 47907, USA
}

\corres{Luca Vaccino}{lvaccino@purdue.edu}

\begin{document}

\maketitle

\pagestyle{empty}

\begin{abstract}
Establishing long-term human settlements in deep space presents significant challenges. Harsh environmental conditions, such as extreme temperature fluctuations, micrometeorite impacts, seismic activity, and exposure to solar and cosmic radiation pose obstacles to the design and operation of habitat systems. Prolonged mission duration and the vast distances from Earth introduce further complications in the form of delayed communication and limited resources, making autonomy especially desirable.

Enabling simulation of the consequences of disruptions and their propagation through the various habitat subsystems is important for the development of autonomous and resilient space habitats. While existing simulation tools can assist in modeling some of these aspects, the integration of damage propagation, detection and repair in a computational model is rarely considered. This paper introduces and demonstrates a simulation architecture designed to model these aspects efficiently. By combining physics-based and phenomenological models, our approach balances computational efficiency with model fidelity. Furthermore, by coordinating subsystems operating at different time scales, we achieve real-time simulation capabilities.

After describing the architecture, we demonstrate its application within HabSim, a space habitat system model developed by the NASA-funded Resilient Extraterrestrial Habitat Institute (RETHi). In these scenarios we consider fire hazard propagation within a lunar habitat to illustrate both how our architecture supports the modeling of disruption propagation, detection, and repair in a simulation environment and how the HabSim model can be leveraged for through stochastic simulations to support resilience assessment. The architecture developed herein is efficient and scalable, enabling researchers to gain insight into resilience, autonomy and decision-making.
\end{abstract}



\section{Introduction}
\label{sec1}

Establishing safe and resilient long-term human settlements in deep space presents significant challenges due to harsh environmental conditions, severely limited resources, and substantial communication delays with Earth. In addition to the absence of an oxygenated and pressurized atmosphere, environmental hazards include extreme temperature fluctuations, micrometeorite impacts, abrasive dust, seismic activity, and exposure to solar and cosmic radiation. Furthermore, the need for Earth-independence drives many of habitat design requirements due to the extended duration of missions and vast distances from Earth, which complicate communication and access to resources \citep{hauplikmeusburgernd}, \citep{crusan2017}. Studying the consequences of disruptions on habitat functions under such extreme conditions and establishing methods for their rapid detection, diagnosis, and repair are essential for ensuring the long-term sustainability and autonomous operation of these habitats \citep{kesslernd}.

To address the operational challenges posed by deep space habitats, habitat systems are typically viewed as systems-of-systems (SoS), which are characterized by the presence of independent constituent systems that interact to produce a collective behavior not achievable by individual systems alone \citep{maier1998}. Systems engineering principles have been widely applied in the military \citep{ChenUnewisse}, aerospace, and disaster management sectors \citep{little2019}. Research in SoS focuses on developing models and simulation architectures to support the design and operation of complex systems. These simulated environments are critical for evaluating the effects of disruptions, repair strategies, and autonomous decision-making policies \citep{gracianoneto}. Simulations also provide a means to explore scenarios that may be impractical or too risky to test in real-world settings \citep{baheri2023}. Using multi-fidelity models allows for rapid design iterations and immediate feedback on modifications, a vital aspect of accelerating the design process \citep{maier2014}. Such simulations are essential to the ongoing development of digital twins for complex systems \citep{NAP}, and they offer numerous advantages, including enabling virtual experiments, exploring visual interface designs, and reducing the costs and risks of physical trials \citep{cristaldi2005}.

While simulation is important to study the emergent behavior of complex systems, developing models with sufficient fidelity to accurately describe the interactions between subsystems often either leads to high computational demands or lacks the detail needed for effective design. Achieving a balance between model fidelity and computational efficiency is therefore a significant challenge. In some cases, detailed physics-based models are replaced with phenomenological models that capture system behavior with less computational effort. However, current models provide limited avenues for incorporating faults, their propagation, detection and repair.

A comparison of existing studies, discussed in detail in Section 2, reveals that while physics-based models offer valuable insights into the quantitative progression of events, they are time-consuming and difficult to implement. In contrast, phenomenological methods, which focus on system behavior, often suffer from subjectivity and may fail to accurately represent key quantitative aspects of the system. Therefore, combining the strengths of both modeling approaches is essential to reduce computational demands, enhance scalability, and retain the physically meaningful data necessary for studying and quantifying resilience and supporting autonomous decision-making strategies in complex systems like space habitats.


The goal of this paper is to demonstrate the advantages of integrating phenomenological features into a physics-based architecture for disruption propagation in system of systems modeling. The proposed architecture, developed for use within HabSim \citep{AzimiLundFuEtAl}, supports the simulation and propagation of disturbances, the detection of anomalies, and their subsequent repair across a complex system. The HabSim testbed is now being used to study resilience and autonomy in smart habitats \citep{{cilentothesis2022},{pritchardthesis2023},{ulmerthesis2023},{dyke2021},{dyke2022}}. Herein, we describe our approach and demonstrate its use for propagation of faults within a simulated space habitat, highlighting the architecture’s flexibility in supporting research on resilience-based design and autonomy.

The remainder of this paper is organized as follows: In Section 2, we review and compare different modeling strategies and software architectures from the literature to identify lessons and reveal gaps. In Section 3, we describe our technical approach, with a focus on the architecture. Section 4 presents an illustrative example where we use the architecture developed to study disruption propagation in a lunar space habitat. We demonstrate its integration into HabSim. Using this architecture within the HabSim testbed, we simulate a fire event and its cascading effects, followed by the detection, and suppression of the fire. We then demonstrate how HabSim can be utilized to perform stochastic simulations of disruptions, thereby allowing for the investigation of system resilience under hazards. 


\section{Technical Background}

Herein we discuss relevant background work toward the development of systems modeling and simulation capabilities, and associated architectures adopted. This review motivates the need for developing an efficient approach to enable research that will lead to insights into resilience and autonomy, as well as providing strategies for the development of this novel architecture.

\subsection{Modeling Strategies for Fault Simulation}

In surveying recent studies in systems modeling, we categorize the various approaches into four levels, in terms of their ability to model faults, propagate damage, detect faults, and initiate repairs. Each level also considers both physics-based and phenomenological modeling capabilities. We define the four levels as follows: Level 1 models focus on fault modeling and propagation; Level 2 extends the architecture to include fault detection; Level 3 adds the ability to diagnose detected faults; and Level 4 integrates repairability. Exploring each of these levels enhances our understanding of how faults affect various systems within a complex system.

Examples of Level 1 models are commonly found in mechanical and aerospace engineering research. For instance, \citet{Zhao} model fault propagation in a small aircraft's fuel supply system using interface automata. Interface automata are finite state machines that model interactions between a system and its environment through well-defined interfaces. The advantage of using interface automata lies in the ease of propagating faults across different systems. However, the development of such models can be quite subjective, as different modelers may offer varying reasoning for fault propagation based on their experience. \citet{DiaoEtAl} explore fault scenarios in computer systems, proposing a phenomenological model for fault generation and propagation with a focus on hardware and software fault interactions. While this model offers advantages for studying fault propagation, it does not consider the temporal aspect of fault propagation, limiting its capacity for fault detection and repair strategies. Additionally, \citet{Dibowski2016FaultPropagation} combine an ontology-based building information model with rules encoding expert knowledge to model faults in building automation systems. \citet{Vaccino} examine fault modeling in an axial piston pump using physics-based modeling. This model provides a high level of fidelity, allowing differentiation between various faults, but at the cost of significant computational demands. The study by \cite{Plotner2013} presents a mathematical model of air leakage; however, it does not address the impact of such failures on surrounding habitat systems and components. \citet{ChenUnewisse} conduct multi-physics modeling of solid rock matrices fully saturated with a single fluid phase, modeling mechanical, hydrological, thermal, and chemical processes to create a detailed fault propagation model, though at a higher computational cost. \citet{Mohammed2020GearFault} model faults in gears using mathematical equations, highlighting the computational cost of finite element method models. \citet{Czupalla2015} simulate a plant culture failure in a Mars habitat and report pronounced phenomenological effects on cabin \ce{CO2} and \ce{O2} levels. While their study employs physics-based models to simulate the habitat’s environmental control system, it does not address fault detection or recovery mechanisms. These examples, along with others in the literature, show that while physics-based fault models tend to be computationally expensive and accurate, phenomenological models, such as the ontology-based models, are faster but require expert knowledge \citep{Hinkelmann2018}.

Level 2 models extend the concepts explored in Level 1 models to include fault detection within different subsystems. For example, \citet{LeonEtAl} introduce a novel method based on the Labeled Transition Systems approach to diagnose multiple faults simultaneously. They developed a tool, DADDY (Distributed Analysis for Distributed Discrete Systems), capable of identifying faults in complex systems. This purely qualitative approach does not employ a model for validation. Similarly,  \citet{Pencole2006Decentralised} adopts a phenomenological approach by proposing a framework where faults are modeled using automata-based interfaces and detected using a decentralized detection algorithm. In this case, the model remains qualitative and does not account for any physical quantities. On the other hand, \citet{li2020} uses a mathematical model to verify a fault detection and discrimination algorithm based on integrated system failure analysis and aims to develop fault diagnosis strategies during the design phase. \citet{WangKarimi} expand an existing chemical process simulation model to implement fault propagation and detection. Examples of Level 2 systems models confirm the trade-off in accuracy and efficiency between physics-based and phenomenological models. The phenomenological approaches reviewed here offer scalability due to their generality, whereas physics-based approaches tend to be more accurate, though limited data availability can challenge model calibration.

Level 3 models add fault diagnosis capabilities to fault propagation and detection in systems simulation. \citet{EkanayakeEtAl} provide a comprehensive comparison of various graphical model-based simulations for fault diagnosis and prognosis in discrete event systems. Petri nets are one example of these models, providing a graphical paradigm for formally describing logical interactions and activity flow within complex systems. Their research demonstrates that while graphical model-based simulations are valuable tools for quickly evaluating fault propagation and detection, constructing such models can be subjective. \citet{lafortune2001failure} use an automata modeling approach based on a discrete model of a system to test failure diagnostic algorithms on a small mechanical system composed of a pump, valve, and controller. These automata represent the system’s behavior as a phenomenological set of states and transitions, enabling systematic analysis and verification. While the model provides an in-depth exposition of state progression, it is limited by the absence of a time component, as it is purely based on discrete event systems. Another study by \citet{Strangas} addresses fault modeling, diagnostics, and prognostics for electrical drives. This research uses methods such as Kalman filters and Markov models and emphasizes the need for experimental data to train fault detection and diagnosis (FDD) models \citep{Strangas}. \citet{LiuEtAl} examine physics-based fault modeling, detection, and prognosis in hydraulic systems, such as hydrostatic bearings, stressing the importance of real data and the difficulty in merging knowledge from different experts. These examples highlight the need for accurate data to develop physics-based models, which are not always available. Conversely, while phenomenological models like automata or Petri nets are easier to generate from user experience, they may be limited in expressing the physical system and can therefore be challenging to apply to practical design and automation challenges.

Level 4 models integrate repairability into systems simulation. \citet{Kumar2019Reliability} use semi-Markov processes to model both fault occurrence and repair times, applying this method to a two-pump standby system to demonstrate enhanced repairability. This method effectively studies dependencies within the model components but lacks information about physical quantities. Another study by the same author utilizes repairable fault trees to understand fault propagation and repairability within systems, helping design or modify systems for more efficient repairs \citep{KumarSharma}. Both studies use binary fault modeling to enhance computational efficiency but are limited in their ability to describe system degradation over time, which is crucial for understanding system resilience. In contrast, \citet{little2019} employ a conceptual framework to study system resilience against coastal flooding. This framework captures core phenomenological behaviors but lacks sufficient data to address the quantitative aspects of various subsystems.  \citet{Alzalab2020FaultRecovery} and \citet{BobbioRaiteri} model fault recovery, detection, and repair in discrete event systems using lumped Markov chain models, with applications in industrial and other systems. Like Little et al. (2019), these models clearly identify fault propagation and recovery across subsystems, but lack any quantitative evaluation a  subsystems.

This discussion of various tools reveals that while physics-based simulation approaches provide more detailed information on the quantitative progression of the system, they are time-consuming and challenging to implement compared to phenomenological methods. In contrast, phenomenological methods have the potential to introduce subjectivity and may fail to represent key quantitative details of the system. All of these factors must be weighed when choosing an approach. A concise summary of the studies and their classification is provided in Table \ref{tab:modelling}.

\begin{table}[H]
\centering
\caption{Summary and classification of the different modelling levels in the literature}

\label{tab:modelling}
\scriptsize
\renewcommand{\arraystretch}{1.5} 
\begin{tabular}{|l|c|c|c|c|c|c|}
\hline
\textbf{Reference} & \textbf{Level 1} & \textbf{Level 2} & \textbf{Level 3} & \textbf{Level 4} & \textbf{Physics-based} & \textbf{Phenomenological} \\
\hline
\cite{Zhao}           & \checkmark &  &  &  &  & \checkmark \\
\hline
\cite{DiaoEtAl}      & \checkmark &  &  &  &  & \checkmark \\
\hline
\cite{Dibowski2016FaultPropagation} & \checkmark &  &  &  &  & \checkmark \\
\hline
\cite{Vaccino}                     & \checkmark &  &  &  & \checkmark &  \\
\hline
\cite{ChenUnewisse}      & \checkmark &  &  &  & \checkmark &  \\
\hline
\cite{Mohammed2020GearFault}      & \checkmark &  &  &  & \checkmark & 
\\
\hline
\cite{Plotner2013}      & \checkmark &  &  &  & \checkmark & 
\\
\hline
\cite{Czupalla2015}      & \checkmark &  &  & & \checkmark & \checkmark \\
\hline
\cite{LeonEtAl}   & \checkmark & \checkmark &  &  &  & \checkmark \\
\hline
\cite{li2022propagation}                & \checkmark & \checkmark &  &  & \checkmark &  \\
\hline
\cite{WangKarimi}          & \checkmark & \checkmark &  &  &  &  \\
\hline
\cite{Pencole2006Decentralised}     & \checkmark & \checkmark &  &  &  & \checkmark \\
\hline
\cite{LiuEtAl}        & \checkmark & \checkmark &  &  & \checkmark &  \\
\hline
\cite{EkanayakeEtAl} & \checkmark & \checkmark & \checkmark &  &  & \checkmark \\
\hline
\cite{Strangas}  & \checkmark & \checkmark & \checkmark &  & \checkmark &  \\
\hline
\cite{BobbioRaiteri}    & \checkmark & \checkmark & \checkmark &  &  & \checkmark \\
\hline
\cite{KumarSharma}        & \checkmark & \checkmark & \checkmark & \checkmark &  & \checkmark \\
\hline
\cite{BuddeEtAl}    & \checkmark & \checkmark & \checkmark & \checkmark & \checkmark &  \\
\hline
\cite{little2019}                   & \checkmark & \checkmark & \checkmark & \checkmark &  & \checkmark \\
\hline
\cite{Alzalab2020FaultRecovery}    & \checkmark & \checkmark & \checkmark & \checkmark &  & \checkmark \\ \hline

\end{tabular}
\end{table}

This review of existing studies illustrates that very few studies include all four levels. Further, physics-based models provide precise quantitative insights into a system's evolution, but are often computationally demanding to implement. Conversely, phenomenological approaches focus on system behavior but can be subjective and may lack accuracy in representing key quantitative aspects. To address these challenges, an approach that combines the strengths of both approaches is appropriate. Thus, we adopt a hybrid modeling approach that not only reduces computational demands and enhances scalability, but also maintains the physical accuracy necessary in key subsystems for examining resilience-based design methods and developing and evaluating methods for autonomous decision-making. 

\subsection{Overview of Software Architecture Patterns}

 Herein, we discuss different software architectures and architectural modelling patterns proposed in the literature, consider their strengths and drawbacks, and adapt these methods for the development of our modular architecture.

 Architectural patterns aim to establish reusable architectures for SoS, and offer solutions to coordinate different systems. In particular, \citet{Sanduka2014ModelBased} compare different modeling patterns and discuss the importance of using time-triggered and event-triggered events to meet real-time requirements.\citet{Klein2013Systematic} propose a set of available architectural patterns and emphasize the need of a centralized decision-making authority to assess reliability and resilience. Different type of signals are usually used to connect subsystems in a complex system, and a mediator is required to manage such a complex network which enables interactions between the heterogeneous entities. Different software mediators types are discussed by \citet{Garces2018}. They show how the router mediator can be used to transmit data between different entities, and show the necessity of a data filter to transmit heterogeneous packages of data to different systems. This ability is particularly useful for systems exchanging different data types, and can be applied to our architecture to combine signals from phenomenological and physics-based models.

Implementation of a mediator facilitates the integration of different signals and subsystems. However, additional requirements need to be defined to ensure the feasibility of real-time simulation. In particular, we focus on three key aspects of available architectures. The first aspect is scalability, the ability of a software to accommodate growth \citep{gorton2022} and effectively adjust to changing demands in user load and processing volume while maintaining computational efficiency \citep{Lucidchart2022Scalability}, \citep{bondi2000characteristics}.
The second aspect is agility, which concerns the speed at which software can respond to evolving design requirements \citep{AgileManifesto} \citep{Laanti2013Agile}.
The third aspect is responsiveness, which measures the system's promptness in completing tasks within a specified timeframe, ensuring tasks are executed efficiently within the designated period \citep{anderson2009responsiveness}.

A review of several fundamental software architecture patterns \citep{Richards2022Software} reveals that a focus on the event-driven and  microkernel architectures is warranted due to their relative capacities for scalability, agility, and responsiveness. Event-driven architectures (EDA) comprise key components such as the event producer, mediator, and event consumer. Events serve as triggers within the architecture. EDA can operate in two modes: synchronously, where the mediator aligns incoming requests with outgoing event signals, and asynchronously, where timing is determined by the event consumer's schedule \citep{Chandy2009EventDriven}. EDA is highly agile, as components can be incorporated or altered in response to evolving requirements, all while avoiding extensive refactoring \citep{Michelson2006EventDriven}. This architecture is also scalable, allowing the integration of additional processing units and services to manage augmented workloads and broaden functionality \citep{Richards2022Software}. Finally, EDAs typically demonstrate real-time responsiveness, with system components engineered to promptly react to events \citep{Michelson2006EventDriven}.

Microkernel architecture, by contrast, offers a more structured approach to provide agility \citep{Richards2022Software}. It is centered around a stable core, the microkernel, with adaptability provided through plug-ins or modules \citep{RanaBaul2023Microkernel}. This architecture's scalability is facilitated by the ability to add or update these modules without modifying the core logic. Responsiveness in the microkernel architecture is more consistent and predictable, given the stability of the core, although it may not achieve the immediate responsiveness characteristic of EDA \citep{Richards2022Software}. This architecture features a minimalistic core system equipped with essential functions, supplemented by independent plug-in components designed to augment the core system's capabilities.

In summary, EDA is ideal for rapid adaptation with agility and responsiveness, while microkernel offers stable, scalable features suitable for systems needing a consistent core. Combining both can leverage their strengths for specific needs. In the following section, we will discuss how we incorporated the lessons learned from the literature to develop our architecture.
\newline

\section{Disruption Propagation Architecture}

Developing the necessary architecture to support the simulation of fault propagation, detection and repair presents significant challenges due to the overall complexity of the interconnections and integration of the constituent systems \citep{li2020}. The developed system model must have sufficient complexity to capture the fault propagation and recovery process while simultaneously minimizing computational cost for stochastic simulation and other model integration opportunities~\citep{dyke2021} \citep{AzimiLundFuEtAl}. We address these challenges by applying the insights from Section 2 to develop a modular software architecture capable of supporting meaningful, layered interactions between subsystems, all while operating within the constraints of real-time simulation and limited computational resources.

\begin{figure}[h]
\centering
\includegraphics[width=0.4\textwidth]{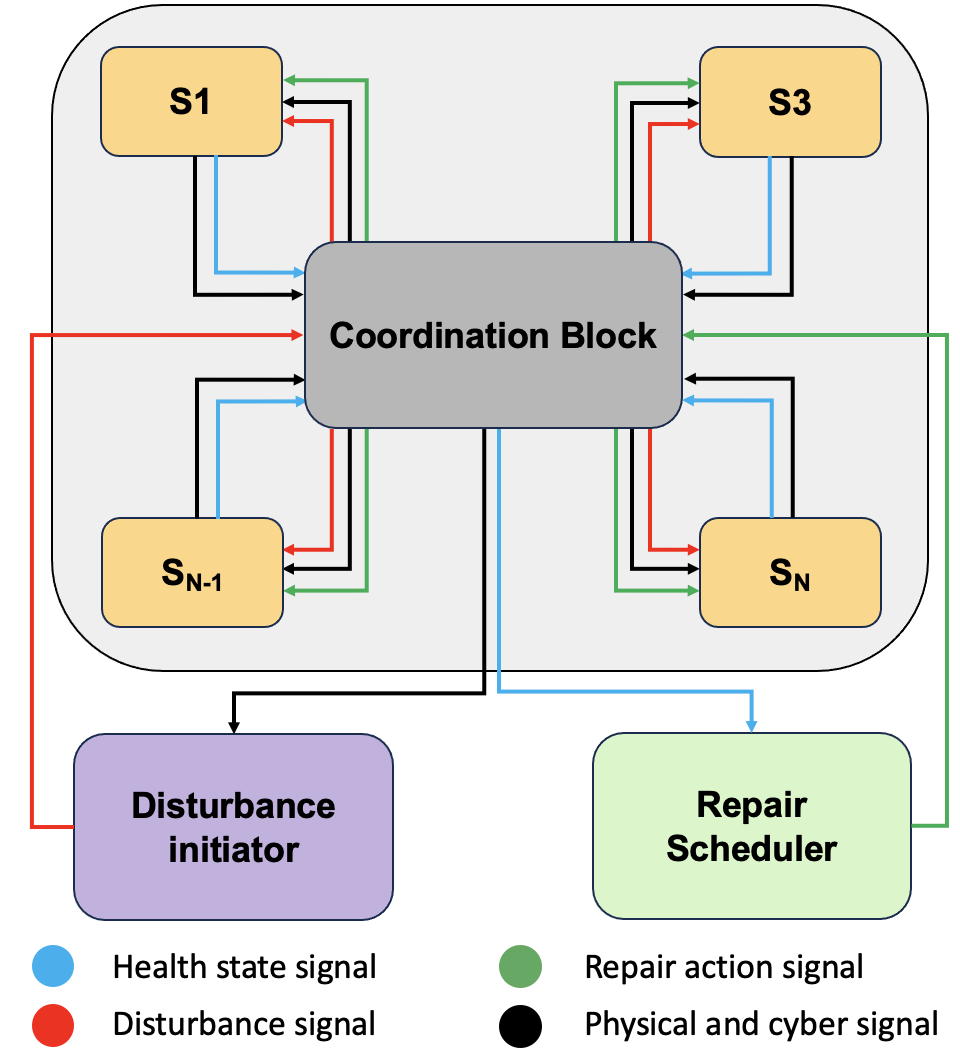}
\caption{Architecture developed for disruption initiation and propagation.}
\label{fig:Framework}
\end{figure}

The scalability of our architecture is improved by standardizing the format of subsystem input and output signals~\citep{sun2022demonstration}~\citep{AzimiLundFuEtAl}.
In particular, in HabSim, we model each system such that it receives a disturbance signal, a repair signal, and a combination of physical signals and cyber signals, thus enabling the subsystems to systematically connect to the disruption initiator system that is the focus of this work.  
Furthermore, we employ the mediator and data filter to manage various signal types, integrating the scalability of the microkernel architecture with the flexibility of the EDA. This approach enhances the overall scalability of the architecture, as illustrated in Fig. \ref{fig:Framework}.

 In Fig. \ref{fig:Framework}, each constituent subsystem of the system is represented in orange, with the purple block represents the disturbance initiator system, and the green block is the repair scheduler system. The disturbance initiator and repair scheduler are modeled using microkernel and EDA architectures. The different systems can be seen as plug-in elements, since they can be added, changed, or removed to model different systems. For example, if more functionalities are to be added in the future, it will be sufficient to add a new system using the standardized input and output block format. Similarities with the EDA lie in the working principles of disruption and repair initiation. A disruption is activated only if the intensity level is greater than one, i.e. when an event occurs, while the repair is initiated only after the fault has been detected by the agent.

 The dataflow between the disturbance initiator system and other nearby systems is depicted in red in Fig. \ref{fig:Framework}. This flow is based on the use of both physics-based and phenomenological signals. The former refers to all quantities that can be measured and used as inputs for mathematical equations, while phenomenological indices are used to model disruption propagation at a higher level and are identified as intensity levels. Using both physical and phenomenological signals combines accuracy and computational efficiency. Physical signals are tangible, measurable, and expressed with units, while phenomenological signals are narrative, indicating an effect to be initiated in the receiving subsystem. Phenomenological signals are primarily used when computational speed is prioritzed, such as in the development of cyber-physical testing regimes or in the stochastic analysis of system resilience; however, not all subsystems require them and may instead rely on physics-based signals when available. 

In this architecture, we aim to support a variety of disruption scenarios, each leading to distinct outcomes and enabling diverse situations. We define the severity of primary disruptions using intensity levels. Then, to model the secondary effects of a disruption, we use damage indicators~\citep{AzimiLundFuEtAl}. For instance, in a fire scenario, the initial disruption effect is defined through an intensity level, which causes the habitat to  increase its temperature. This temperature increase is then communicated from the interior environment to other subsystems through a damage indicator, which allows the effects of the disturbance to cascade to other systems (through damage indicators) leading to a loss of performance of power converters, energy storage and cooling fans. The use of intensity levels and damage indicators will be further clarified in the illustrative example.

The coordination block, shown in gray in Figs.\ref{fig:Framework} and \ref{fig:HabSim_impl}, manages the interactions between all systems at each simulation time step. Within the coordination block, physics-based and phenomenological signals are merged and routed to the appropriate subsystem \citep{AzimiLundFuEtAl}. To handle the various types of signal and their timings, the coordination block directs these signals into an internal data fusion block, which is responsible for aggregating and routing them to the corresponding systems. The concept of a data fusion block, an aggregator, and a router block is introduced in \citep{Garces2018} in the context of a flood detection system, though in this case, the integration of physical signals within the model is not addressed.

When a disruption propagates to a subsystem (Fig. \ref{fig:Framework}), it triggers an FDD test that detects anomalies in sensor data and updates the system health state indicator (blue line in Fig. \ref{fig:Framework}). If the indicator exceeds a user-defined threshold, the repair scheduler issues a repair command to the affected subsystem. The repair signal, illustrated by the green arrows in Fig. \ref{fig:Framework}, consists of physics-based quantities such as the repair rate which can be the speed at which an agent removes dust from solar panels or patches a hole in a habitat layer. These parameters, including the thresholds, are configurable through the simulation input files to accommodate different research objectives~\citep{AzimiLundFuEtAl}.

\subsection{HabSim Implementation}

\begin{figure}[H]
\centering
\includegraphics[width=0.8\textwidth]{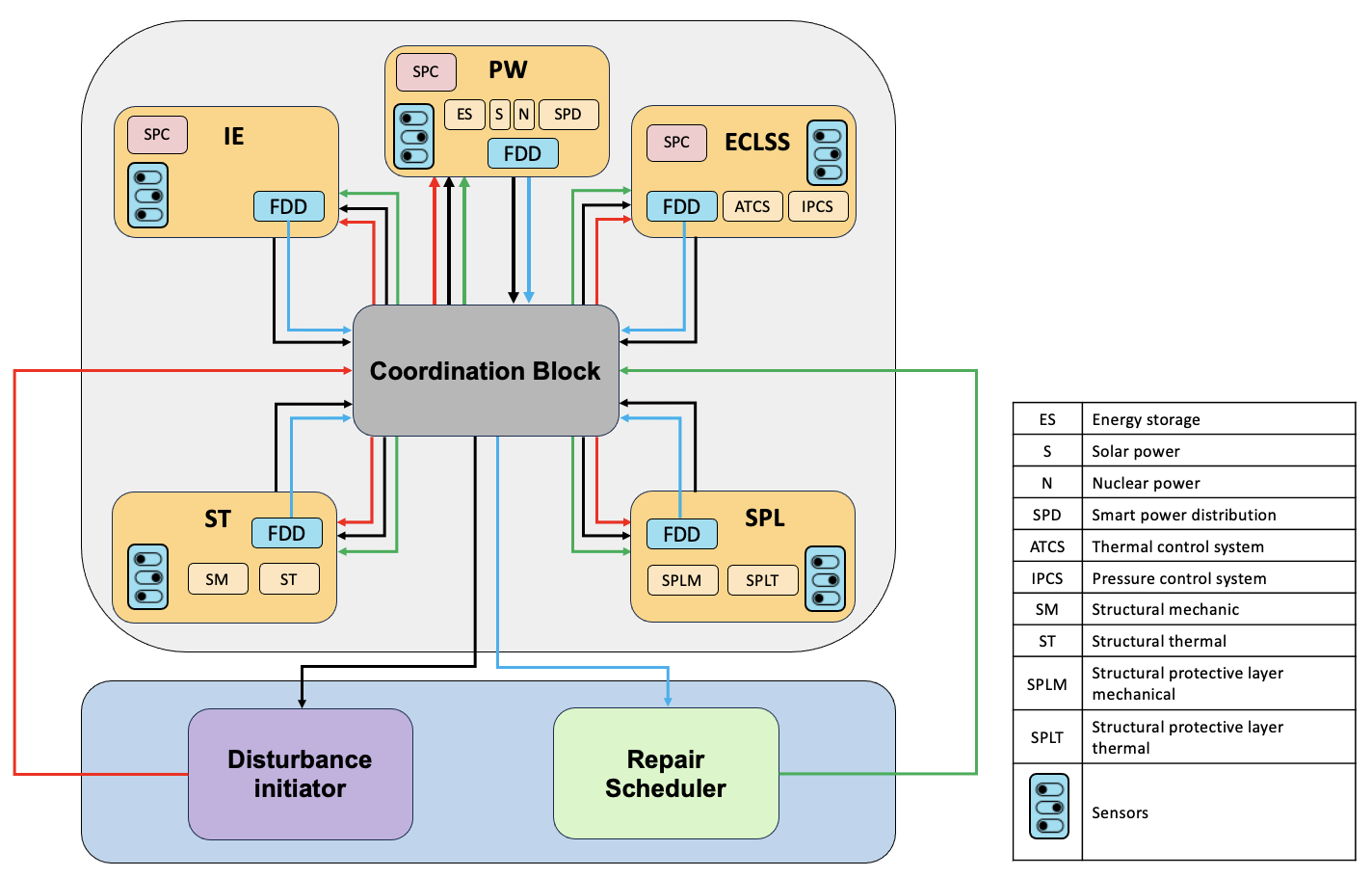}
\caption{HabSim architecture, its component systems, and their communication pathways. Blue represents cyber signals, green indicates repair actions, black denotes physical signals, and red signifies disruptions.}
\label{fig:HabSim_impl}
\end{figure}

Figure \ref{fig:HabSim_impl} illustrates the specific implementation of the proposed architecture in the HabSim testbed~\citep{AzimiLundFuEtAl}. HabSim may be used to expose the habitat system to several disruptions, including fire within the habitat, seismic activity, micrometeorite impact, and lunar dust effects, among others. The effects of those disruptions are modeled using both physics-based and phenomenological signals and are characterized based on environmental models discussed in~\citet{AzimiLundFuEtAl} and in the Habsim documentation written by the \citet{HabSimUsermanual}. 

The system is separated into two distinct components, the nominal state model, enclosed into the gray box, and the disturbance and repair initiation model, enclosed into the blue box in Fig. \ref{fig:HabSim_impl}.
Each of the modeled systems contributes to the physical understanding of the overall system behavior, where specific interactions and fault propagation mechanisms are targeted for phenomenological modeling to add hazard representation and complexity throughout the system while easing subsystem communication and improving real-time capacity.

\begin{figure}[H]
\centering
\includegraphics[width=0.5\textwidth]{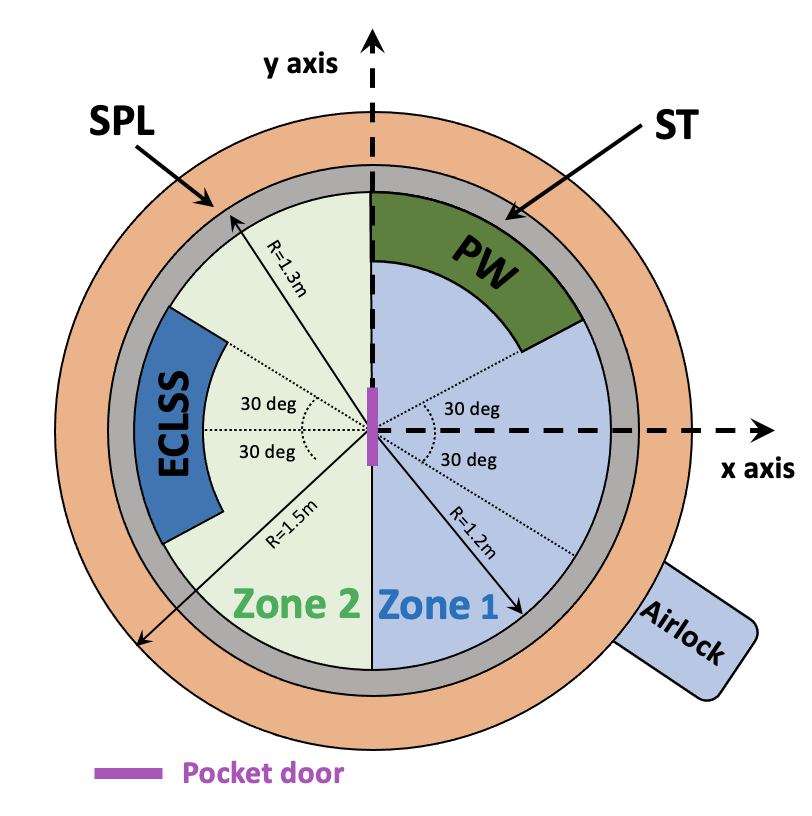}
\caption{Planar view of the modeled habitat.}
\label{fig:HabSim_map}
\end{figure}

The modularity of our architecture enables the substitution of subsystems with models of varying complexity, allowing for adjustments in the fidelity of the results. A standard notation, defined in \citep{AzimiLundFuEtAl}, supports the integration of models with different levels of fidelity within our architecture. Several authors have used HabSim for research into resilience, such as the work by \citep{jain2024control}, and others have substituted their own expanded models into HabSim for specific investigations using this architecture \citep{rhee2023damageable} \citep{chebbo2022microgrids} \citep{chebbo2024fault}\citep{Shahriar2024}. A planar view of the modeled habitat is represented in Fig.  \ref{fig:HabSim_map}, which shows the physical arrangement of the subsystems represented by the orange blocks in Fig. \ref{fig:HabSim_impl}. 

The Interior Environment (IE) model captures the pressure and temperature of the volume of air inside the habitat. The IE model considers two zones, and also includes elements such a simplified airlock interface, which in case of damage can cause an air leak to the external environment. A pocket door that is used to divide the IE volume into zone 1 and zone 2 when necessary. IE supports physics-based interactions with structure (ST) and structural protective layer (SPL), where these interactions are characterized by thermal and pressure dynamics. Phenomenological interactions supported by this subsystem consist of the interactions between the IE and the fire disturbance. The radiation heat modeled in the IE as a result of a fire is represented phenomenologically through five different intensity levels. The habitat ST serves to enclose the IE. It is covered by the SPL, which insulates the interior from the harsh exterior environment conditions, such the extreme temperatures, and also serves as protection from radiation. 

The habitat temperature and pressure are controlled by the environmental life control system (ECLSS). In the current configuration, ECLSS is composed by two primary subsystems: the pressure controller (IPCS) for regulating habitat pressure, and the active thermal control system (ATCS) for temperature management. ATCS activates heaters if the measured temperature is too low, and engages the cooling system when the habitat air temperature is too high. Pressure is managed by introducing air into the habitat when the interior pressure exceeds the lower safe boundary and venting air out when it is too high. Physical interactions occur through the pressure and temperature changes, while phenomenological interactions are used to initiate the ECLSS compressors, heaters, evaporator and fan damage caused by fire, moonquake or micrometeorite impact.

The power system (PW) models solar and nuclear power generators, and energy storage. Within PW, a smart power distribution (SPD) system is also modeled that is able to prioritize power loads for optimal crew safety and consumption \citep{inproceedings}. In this case, the power consumed by ECLSS loads and PW loads or the dust deposition on solar panels are modeled as physics-based interaction, while thermal effects on battery efficiency are modeled phenomenologically.

Each subsystem contains a sensor block which includes all the sensors used to monitor habitat states, such as pressure, temperature, and acceleration. Accelerometers monitor structural health, while temperature and pressure sensors are used for thermal and pressure management. The blue blocks in Fig. 3 represent the synthetic FDD, which identify sensor data deviations from nominal values and, in case of values outside the safe range, it activates the repair action. In HabSim version 6.3, the FDD interprets the health states for each subsystem and reports its assessment, in the form of a binary health state for each subsystem component (0: healthy, 1: damaged), to the repair scheduler block. 
The blue box in the lower section of Fig. \ref{fig:HabSim_impl} conveys the structure of the system damage and repair interactions. The block in purple represents the disturbance initiator, that is the system which initiates disruption into different subsystems. The disturbance initiator models the harsh exterior environment and disruptions like moonquakes, fires, micrometeorite impacts, dust accumulation, airlock leaks, nuclear system coolant leaks, and dust deposition. Within the disturbance initiator, both physics-based and phenomenological interactions are implemented. Meteorite and moonquake intensity levels are example of phenomenological interaction with PW, and ECLSS, causing damage to their component and structures. In case of meteorite impact on ST, hole size is determined by using a phenomenological model.

Note that the disturbance parameters, including intensity level, start time, and finish time, may be provided by the user as discussed in~\citep{AzimiLundFuEtAl}. 
Five intensity levels are used to represent the severity of a disruption, ranging from level one (no disruption) to level five (highest severity and worst outcome). As a disruption propagates, the intensity levels influence subsystems differently depending on their design and vulnerability. This distinction allows the framework to flexibly model a wide range of subsystem behaviors and enhances its applicability across diverse scenarios.
For the start time and completion time, the user also has the option to set the disruption time by sampling values from a probability distribution, allowing for triggering events at random times.

The green block  in Fig. 3 represents a simple agent model, which is triggered by FDD results~\citep{krishnan2024habsim}, and sends information about the repair type, duration, and agent availability to the affected subsystems. Changing those parameters makes it possible to model either human or robot repair action and study the impact of repair time on resilience. 
The result of the described model approach is highly scalable and can be easily improved by adding or updating different blocks.

Once a disturbance has been initiated, the secondary effects of the disruption, which are often obscured in purely physical models, are explicitly captured using damage indicators.

While intensity levels represent the immediate effect of a primary disruption (e.g., fire or micrometeorite) on the system, damage levels model the secondary or cascading effects resulting from that primary disturbance.  These indicators function similarly to intensity levels and use a 5-level scale to represent varying degrees of damage anticipated in a given system. The damage levels are generated in the disturbance initiator and then transmitted to other subsystems, where they act as initiating events for further propagation.

For instance, in the case of a fire, the disturbance initiator sends both the fire phenomenological intensity level and the corresponding physical temperature signal to the coordination block, which then forwards these inputs to the Interior Environment (IE) subsystem. Based on the intensity level, the IE model simulates different heat exchange profiles between the habitat’s air volume and the fire. The resulting temperature influences physical models of pressure and temperature throughout the habitat. The physical habitat temperature is then fed back into the coordination block and routed to the original disturbance initiator. There, it is transformed into a phenomenological damage indicator, which captures the cascading effects of the fire. The resulting damage levels are distributed to impacted subsystems such as PW and ECLSS.

Upon receiving these damage indicators, PW and ECLSS adjust the operational efficiency of their respective components, including compressors, power converters, and energy storage units. For energy storage systems, moderate temperatures (-1 $^\circ$C to 30 $^\circ$C) correspond to minimal damage, while extreme temperatures ($\leq$  -50 $^\circ$C or $\geq$ 50 $^\circ$C) result in maximum degradation. Compressor performance similarly deteriorates outside the optimal range of 10$^\circ$C to 30 $^\circ$C, with extreme damage at $\leq$-20 $^\circ$C or $\geq$ 120 $^\circ$C. Power converters show increasing vulnerability beyond 105 $^\circ$C, with critical failure observed above 145 $^\circ$C.

These classifications allow the subsystem interfaces to efficiently highlight the temperature sensitivity of key system components, providing insight into operational thresholds and resilience under varying thermal conditions.

\begin{table}[H]
\centering
\small
\renewcommand{\arraystretch}{1.5}
\caption{Damage indicator levels for different components and temperature ranges.}
\begin{tabular}{|c|c|c|}
\hline
\textbf{Component} & \textbf{Temperature range [°C]} & \textbf{Damage level [-]} \\
\hline
\multirow{5}{*}{Energy Storage} 
& [-1, 30]            & 1 \\
\cline{2-3}
& [-20, -1]           & 2 \\
\cline{2-3}
& [-20, -50]          & 3 \\
\cline{2-3}
& [30, 50]            & 4 \\
\cline{2-3}
& $\leq -50$, $\geq 50$ & 5 \\
\hline
\multirow{5}{*}{Compressor} 
& [10, 30]            & 1 \\
\cline{2-3}
& [0, 10], [30, 60]   & 2 \\
\cline{2-3}
& [-10, 0], [60, 90]  & 3 \\
\cline{2-3}
& [-20, -10], [90, 120] & 4 \\
\cline{2-3}
& $\leq -20$, $\geq 120$ & 5 \\
\hline
\multirow{4}{*}{Power Converters} 
& $\leq 105$          & 1 \\
\cline{2-3}
& [105, 125]          & 2 \\
\cline{2-3}
& [125, 145]          & 3 \\
\cline{2-3}
& $\geq 145$          & 4 \\
\hline
\end{tabular}
\label{tab:damage_levels}
\end{table}

\section{Illustrative Examples}
Here we present two examples to illustrate how the combination of physical and phenomenological modeling approaches effectively simulates disturbances, their propagation, detection and repair in a habitat system. 
Here we consider the scenario in which a fire is started and ultimately suppressed within the habitat. The results not only demonstrate the development of realistic interactions between habitat subsystems within the simulated habitat, but also highlight the flexibility of the chosen architecture. In the first example a single fire scenario is examined in detail. Detailed information on the numerical results of this scenario on the different habitat subsystems can be found in Appendix A. In the second example realizations of a stochastic simulation are generated to illustrate how the developed architecture may be leveraged to efficiently study system resilience. 

\subsection{Example 1: Fire Disruption Affecting Power System and Interior Environment}
The simulation starts with the system in its nominal condition, with temperature and pressure set to have a safe environment for the human occupants.
At each simulation time step, each system sends physical signals and cyber signals to the coordination block. Physical signals are used as input for mathematical equations, which for example model the heat transfer between the habitat systems. Cyber signals are signals simulating data from sensors, and include sensor noise and uncertainties. Those signals are used as inputs for controllers, like the pressure or temperature control. 
To reduce the computational load and meet real time requirements, each subsystem is constrained to a specific time step for the evaluation of their internal dynamics. 
Capturing the structural vibration using accelerometers is an example of an event with fast dynamics, and it requires a high sampling rate. Events like battery discharge have slower dynamics, therefore the PW system signals can be sampled at a lower sampling rate.
From a physical point of view, the initial conditions for this simulation are the PW energy storage charged at 80{\%}, and the nuclear and solar power generation fully available. 
The temperature and pressure are within the safe ranges for crewed operations, where mostly of the power consumption is due to the temperature and pressure control through the ECLSS subsystem and FDD sensor operations. In this simulation, scientific instruments, housekeeping and rover charging loads have been deactivated to focus on the effect of fire disruption on ECLSS thermal and pressure power consumption. 
Since the power draw is lower than the power generated, the PW energy storage is increasing. The pocket door dividing the two zones is kept open for the entire simulation.
   
\begin{figure}[H]
\centering
\includegraphics[width=0.95\textwidth]{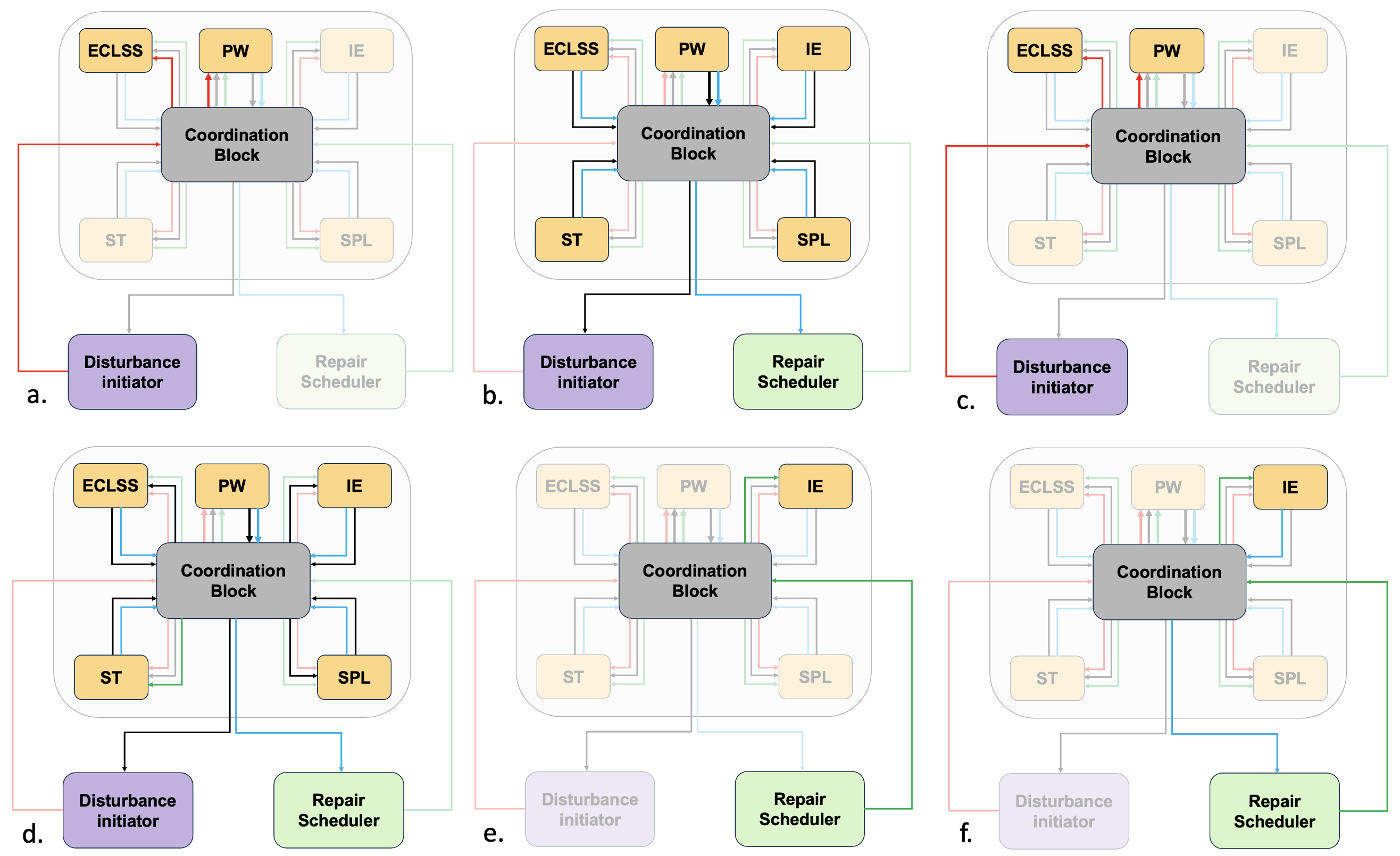}
\caption{Steps through the scenario stages, illustrating disruption propagation, detection, and repair.}
\label{fig:DisrProp}
\end{figure}

\begin{figure}[H]
\centering
\includegraphics[width=0.95\textwidth]{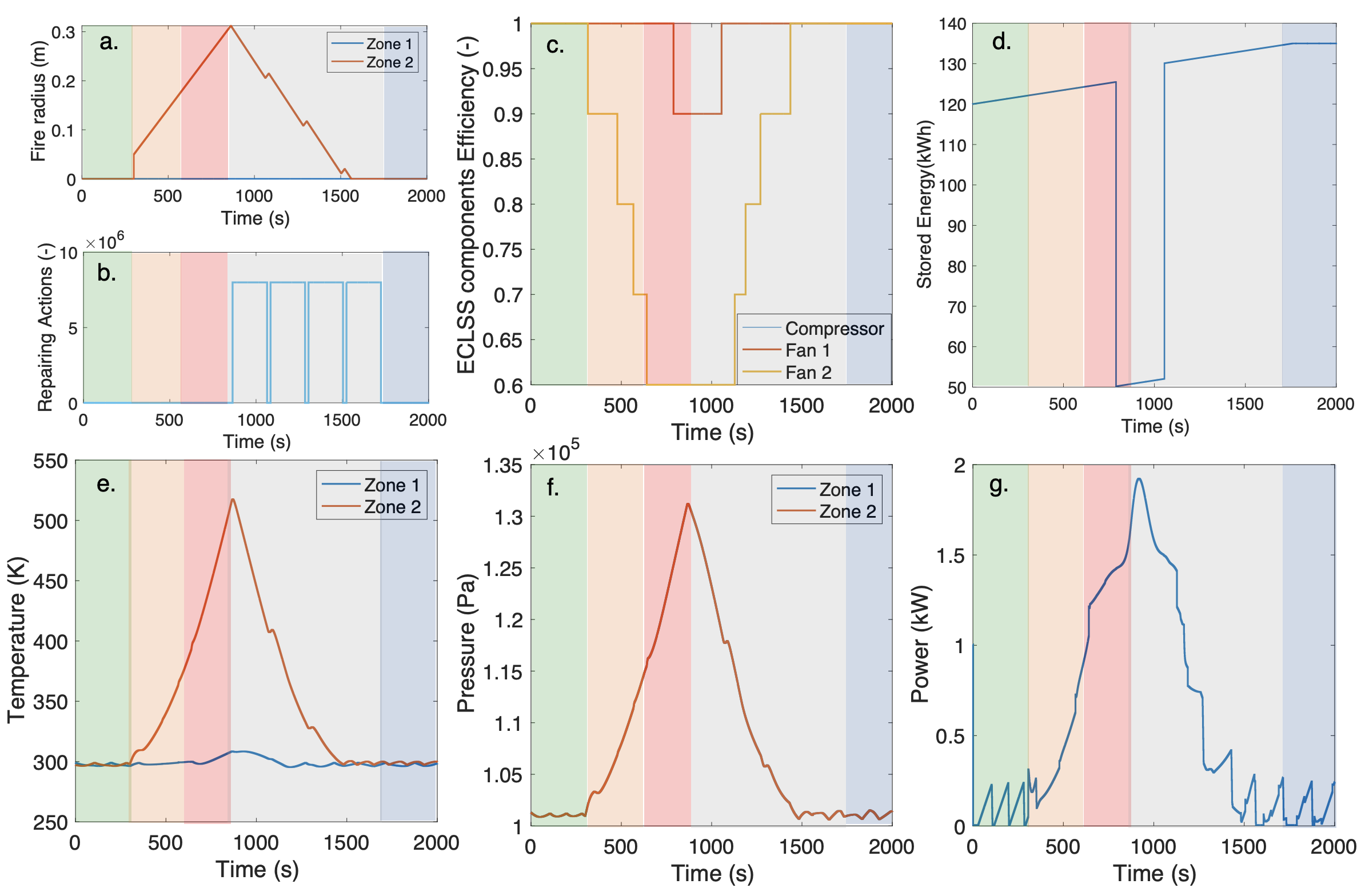}
\caption{Visualization of key parameters during a fire scenario, including fire radius (a), fire suppression steps (b), ECLSS component efficiency (c), stored energy levels (d), temperature in Zone 1 and Zone 2 (e), pressure in Zone 1 and Zone 2 (f), and power consumption (g). Each color represents a different stage of fault propagation, detection, and repair. A more detailed breakdown is provided in Appendix A.}
\label{fig:FireData}
\end{figure}

After running under nominal conditions for 300 seconds, a fire is generated by the disturbance initiator, shown in purple in Fig. \ref{fig:DisrProp} (a). Information about the fire is generated in both physical and phenomenological contexts to support cost-effective communication to the subsystems. The physical information consists of the fire radius, spread rate, and location, while the phenomenological information consists of the fire intensity level. The fire intensity level serves as a granular indicator of internal system damage, providing critical input for the PW, IE, and ECLSS subsystems to initiate their respective responses. Meanwhile, the physical fire data is transmitted to the IE, which experiences a corresponding temperature increase, as shown in Fig. \ref{fig:FireData}.
Physics-based signals are used to model the fire behavior, and these signals are subsequently relayed to the coordination block. In this simulation, a fire is modeled in habitat zone 2. It starts at location [0,-0.75] m which does not immediately damage any of the internal electrical components (e.g. ECLSS and PW subsystems). The fire intensity level is set to 3, which is indicative of a plastic or polymer fire \citep{carmignani2018correlation}, with a spread rate of 0.4 mm/sec. This spread rate has been scaled with respect to the one fifth scale HabSim model \citep{AzimiLundFuEtAl}.

Once the disturbance signal has been initiated and passed to the coordination block, the information is then routed to all systems that are affected by such a disruption, which in this case includes the  ECLSS, IE, and PW subsystems (Fig. \ref{fig:DisrProp} (a)). Initially, information sent to the ECLSS and PW subsystems indicates no hazard, or fire intensity level 1. Internal tracking within the disturbance initiator continuously sends updated information to the appropriate subsystems as the fire spreads at a speed of 0.23 mm/sec. It is important to remember that this value is scaled, since we are considering a one-fifth scaled version of the habitat, and that the fire propagation model is a semisphere propagation model.

Simultaneously, updated states, phenomenological indicators, and sensor readings are routed through the coordination block to the relevant subsystems. In this case study, the evolving temperature within the IE subsystem is sent to the disturbance initiator, which facilitates the propagation of faults from one subsystem to others. Specifically, the disturbance initiator converts the physical temperature information from the IE into a phenomenological damage indicator. This indicator is then transmitted back through the coordination block and used by the ECLSS compressor, and PW power converters and batteries as an indicator of their susceptibility to temperature-induced degradation, as illustrated by the red arrows in Fig. \ref{fig:DisrProp} (c). The correlation between temperature and component degradation is detailed in Table \ref{tab:damage_levels}.
   
This phenomenological approach explores the  high-level temperature effects on system components rather than physical equations, with the intent of developing the complex emergent behaviors between subsystems rather than expressing a comprehensive physical understanding of the damage to each component. Though a purely physics-based modeling approach would develop equivalent emergent behaviors within the system, the increase in computational cost renders the model ineffective for the purpose of preliminary design for resilience and automation. 

From the physical and quantitative perspectives of the simulation, the fire causes the temperature in zone 2 to rapidly increase and it exceeds the safe ranges for crew at time equal to 350 seconds. The temperature increase in zone 2 results in the efficiency decrease of fan and compressor, as can be seen in Fig. \ref{fig:FireData}(c). Because a phenomenological approach is used, the efficiency drop occurs rapidly. Unlike a physics-based model that captures gradual degradation, the threshold-based damage levels lead to step-wise system responses. As a result, transient behavior is not accurately represented, though the steady-state response remains a valid reflection of habitat behavior.
Temperature in zone 1 is affected by the fire as well, because the pocket door dividing the two zones is kept open. However, the temperature in zone 1 increases at a slower rate, and the fan in this zone is still working with maximum efficiency.
Pressure in both zones is the same, because the pocket door is open, and therefore the two zones are in contact, and the IE can be considered as a single volume of air for this study.

The effect of temperature increase and efficiency reduction are reflected on the power consumed by ECLSS cooling system and pressure control, shown in Fig. \ref{fig:FireData} (f). In particular, whenever the fan efficiency decreases, more power is required to keep the fan running at the same target speed. At time equal to 600 seconds, the temperature in zone 2 causes a further reduction on fan and compressor efficiency. As a result, the power requirement from the ECLSS continues to increase. However, as the power generated by solar and nuclear systems is still higher than the consumed power, the the ES power shown in Fig. \ref{fig:FireData} (d) is able to keep up with and exceed the increased demand from ECLSS. 

At time equal to 900 seconds, the high temperature in zone 1 causes a 60 {\%} PW ES efficiency decrease. After that, at time equal to 950 seconds, the fire affects ECLSS condenser resulting in a further increase in power consumption to make up for the lost condenser efficiency. Generated power from solar and nuclear sources still exceeds ECLSS consumption however, and the ES power supply shows a positive slope. 

FDD indicators from each of the habitat subsystems are updated as the hazard progresses and sent to the coordination block at each time step to detect emerging faults (shown as blue arrows in Fig. \ref{fig:DisrProp} (b)). The coordination block synchronizes the FDD signals coming from different systems and routes them to the repair initiator. The fire is detected for this scenario at approximately 850 seconds, when the FDD block identifies an anomaly inside the interior environment system by analyzing the data coming from temperature and pressure sensors. When this anomaly is detected, the health state of the system is adjusted to "damaged" and the appropriate signal is routed through the coordination block to the repair scheduler. Given this information, the repair scheduler initiates the repair action within the IE through the coordination block, where the signal is sent in terms of agent availability, repair rate, and repair type (Fig \ref{fig:DisrProp} (e)).  The agent starts suppressing the fire at time equal to 900 seconds. As soon as the fire starts to be extinguished, the pressure and temperature start to decrease as well. The ECLSS system has been repaired, and the compressor and fan efficiencies start to increase again. 
At time equal to 1100 seconds, the PW ES efficiency is restored to the original value. The power consumption, shown in Fig. \ref{fig:FireData} (g), decreases as soon as the fire start to be extinguished. This is because the ECLSS component efficiency has being restored, so less power is necessary to power the components in nominal conditions. 

The repair action continues until the fire is extinguished, as determined by the health state measured by the FDD. Once the fire is suppressed, the repair scheduler reverts to its nominal condition and continues to wait  for FDD health state signal changes. The simulation then concludes with the pressure and temperature being reestablished within the safe crew ranges, and the ECLSS consumption is back to the nominal conditions. At time equal to 1700 seconds, the PW ES are completely charged, and all the ECLSS components are back to operating in their nominal condition.

In this illustrative example we successfully integrated physical and phenomenological modeling approaches to simulate the effects of fire disruption on a space habitat. We show both primary fire effects (temperature increase) and secondary effects (decrease of PW and ECLSS components efficiency).

\subsection{Example 2: Utilizing HabSim to Analyze Space Habitat Resilience to Fire Disruptions}
The unique disruption architecture of the HabSim allows for the rapid characterization of system resilience under a variety of hazards. While the augmentation of phenomenological components within the physics-based simulations does not guarantee a high-fidelity simulation of the granular damage within the habitat, it does allow for the development of clear insights into critical system vulnerabilities by sampling against hundreds of design, disruption, and recovery variables twice as fast as real-time simulations (using a Dell OptiPlex 5700 PC, Intel64 Family 6 processor). 
In this example, we demonstrate the use of HabSim to evaluate resilience in a space habitat, focusing specifically on assessing one of the parameters of the control effectiveness of a fire suppression system. \cite{jain2024control} defines control effectiveness by four key factors: the availability of a safety control when needed, the likelihood that its design effectively mitigates the targeted hazard, the probability that it will be implemented as intended, and the speed at which it can counteract the hazard before it spreads. The final factor, known as the response margin, measures the time available between fire detection and the moment the fire begins causing permanent damage to habitat components. The response margin is helpful in determining whether the suppression system can respond quickly enough to contain the fire and prevent catastrophic failure.

To simulate fire propagation, we take the fire spread rate as a beta-distributed random variable over the interval [0.23, 1.9] millimeters per second, with shape parameters \(\alpha = 8.49\) and \(\beta = 7.84\). This choice reflects the assumption that the material properties at the fire’s origin do not vary widely, resulting in a relatively constrained range of spread rates, with the beta distribution representing slight asymmetries in the data. The detection time is modeled as a random variable $D$, uniformly distributed over the interval $[280, 560]$ seconds. The total time from the initiation of the fire to its detection is given by $T + D$, where $T$ denotes the time required for the fire to grow to a detectable level, defined as the point at which the fire radius reaches at least 20\,mm.

\begin{figure}[H]
    \centering
    \includegraphics[width=0.7\linewidth]{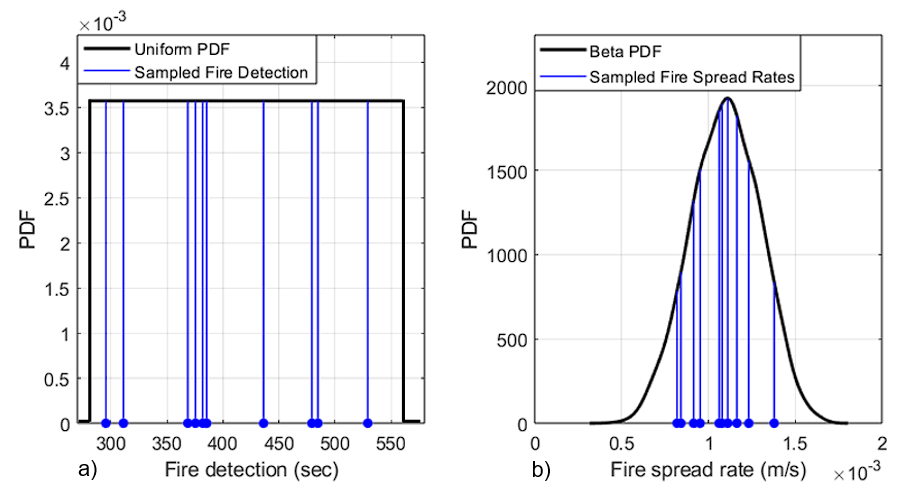}
    \caption{Selected Fire detection samples from uniform distribution (a), and selected fire spread rate samples from beta distribution (b).}
    \label{fig:Distributions}
\end{figure}

The samples of fire detection rate and fire spread used in this study are marked in blue in Fig. \ref{fig:Distributions} (a). Fig. \ref{fig:Distributions} (b) illustrates the fire spread, which are marked in blue as well. These modeling choices are based on the findings of \citet{jain2024control}, who emphasize the importance of probabilistic modeling in evaluating control strategies for space habitat resilience. Their work highlights how stochastic representations of environmental disturbances, such as fire propagation, enhance the robustness of autonomous decision-making frameworks. By incorporating these probabilistic models, HabSim enables a detailed assessment of how different fire spread rates and detection times affect the habitat’s recovery time, internal temperature, and power consumption. These simulations revealed critical thresholds, beyond which fire propagation becomes uncontrollable, and highlighted the importance of early detection and quick response to avoid irreversible damage.
By using the time response margin as a key performance indicator, we support researchers to quantify whether the suppression system can intervene within the critical time window. This approach is fundamental for developing adaptive methods that optimize response actions and improve overall habitat resilience against fire disruptions, and pinpoint the most vulnerable aspects of the habitat design, making it more versatile than traditional, compartmentalized modeling tools.

   


\begin{figure}[H]
    \centering
    \includegraphics[width=0.5\linewidth]{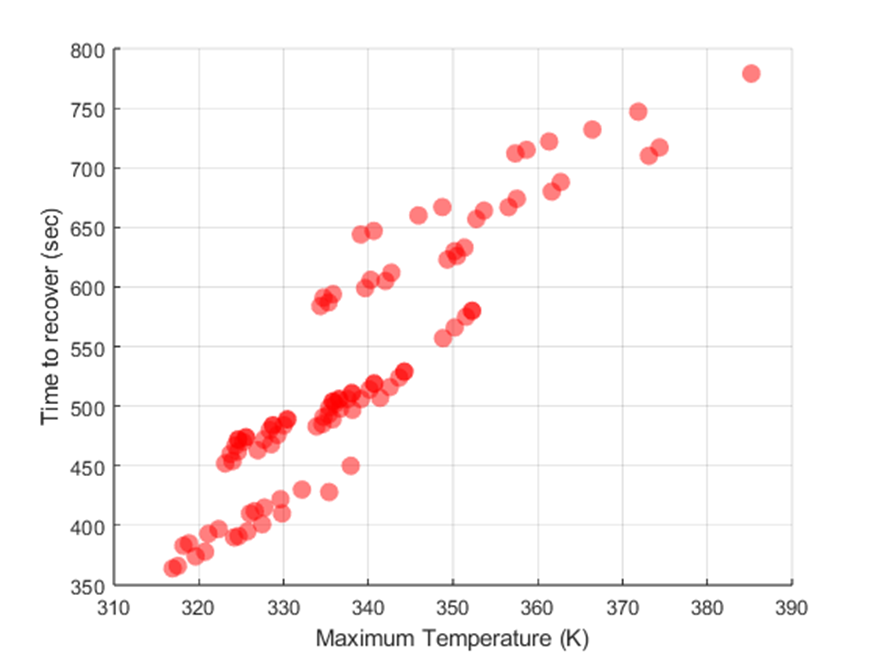}
    \caption{Time to recover and maximum temperature for each simulation.}
    \label{fig:maxT}
\end{figure}

\begin{figure}[H]
    \centering
    \includegraphics[width=0.5\linewidth]{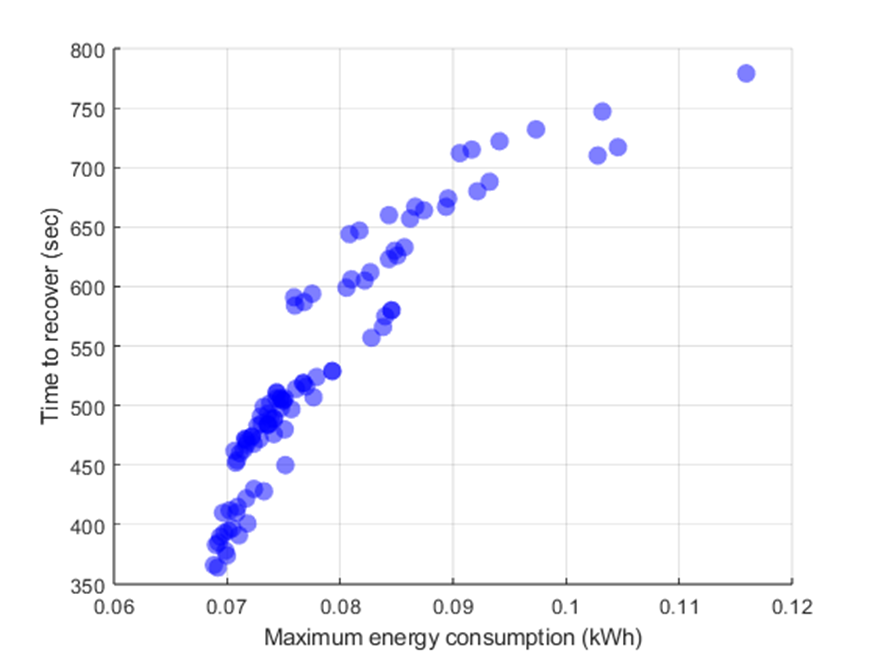}
    \caption{Time to recover and maximum energy consumption for each simulation}
    \label{fig:maxPW}
\end{figure}

After drawing ten samples for the fire spread rate and ten samples for the detection time, the HabSim model is run for every possible combination of these values—resulting in 100 realizations. Ten samples were chosen to balance the need for capturing variability in the system while keeping the computational time manageable. Figure \ref{fig:maxT} illustrates the relationship between recovery time (the time until the fire is extinguished) and temperature, while Fig. \ref{fig:maxPW} presents the correlation between recovery time and habitat power consumption.

The findings indicate that, up to 350 K, recovery time increases significantly with any rise in temperature, highlighting the critical impact of early detection efforts. However, after 350 K, further increases in temperature do not result in a similarly significant increase in recovery time. These results suggests that the temperature is rising at a faster pace, leaving less time before the habitat components are irreversible damaged by the fire. Consequently, the rapid temperature increase also leads to a sharp rise in energy consumption.

\begin{figure}[H]
    \centering
    \includegraphics[width=0.5\linewidth]{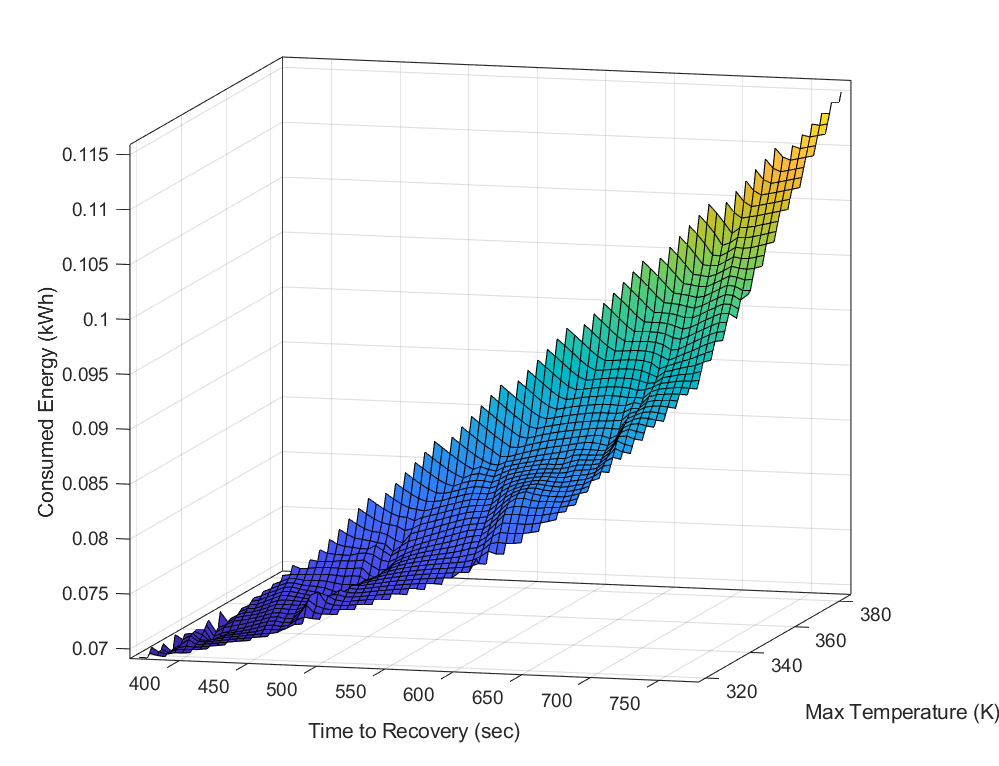}
   \caption{Comparison of time to recovery, temperature and energy consumption for different simulations.}
    \label{fig:3Dplot}
\end{figure}

As illustrated in the 3D plot in \ref{fig:3Dplot}, which collectively examines power consumption, maximum temperature, and recovery time, early detection and prompt suppression of the fire lead to reduced power usage and lower peak temperatures. In contrast, when the temperature nears 350 K, a marked increase in both the temperature and power consumption is observed.


These results can be used to evaluate the effectiveness of control measures and to guide habitat designers in identifying ways to help prevent the rapid propagation of fire hazards. We use the concept of the time response margin, introduced in \cite{jain2024control}. The time response margin quantifies the relationship between the time required for the safety control to mitigate the hazard ($t_{sc, affect}$) and the time it takes for the hazard to spread and potentially lead to dangerous or accident states ($t_{h, effect}$). In the case of a fire, $t_{sc, affect}$ represents the time required to extinguish the fire (time-to-recover), while $t_{h, effect}$ denotes the time at which the fire causes non-recoverable damage (i.e., the time to reach 350 K, after which the time to recover does not experience as significant of an increase with additional temperature).  The response margin is defined as follows:

\[
M_{\text{response}} = 1 - \frac{t_{sc, affect}}{t_{h, effect}}
\]

A higher $M_{\text{response}}$ value indicates more effective safety control. If $t_{sc, affect}$ is less than $t_{h, effect}$, the safety control successfully addresses the hazard before it propagates to other systems. Conversely, if $M_{\text{response}}$ is smaller than one, it implies that the control measure is too slow to prevent the hazard from spreading. Notably, passive safety controls—such as using nonflammable materials that inherently eliminate hazards—achieve a response margin of 1, since $t_{sc, affect}$ is effectively zero.

\begin{figure}[H]
    \centering
    \includegraphics[width=0.7\linewidth]{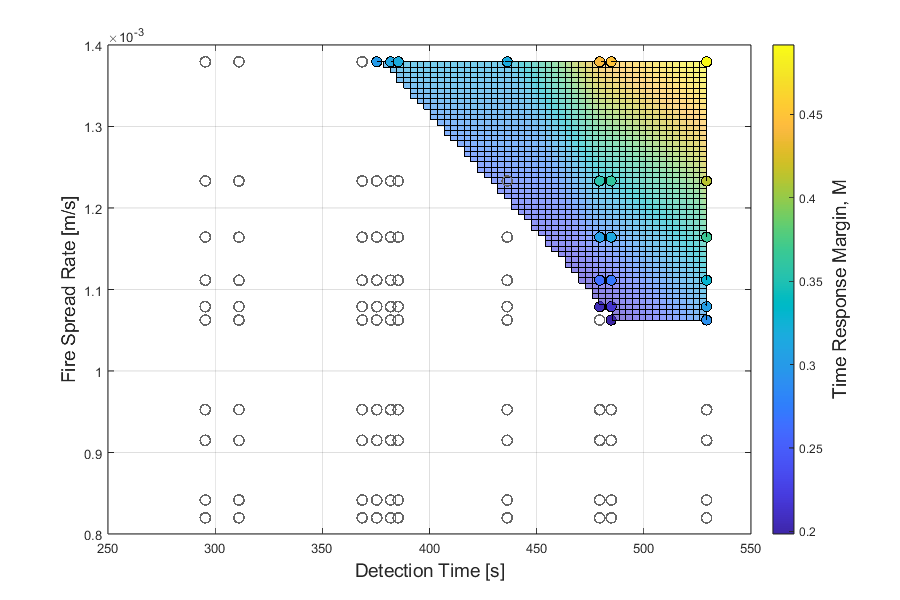}
    \caption{Time to response margin for each simulation.}
    \label{fig:M}
\end{figure}

Figure~\ref{fig:M} presents the results, where an $M_{\text{response}}$ value of 1 signifies that the habitat components remain failure-free. Failures, noted in the plot as values less than 1,  occur when the habitat components sustain irreversible damage. The time response margin starts becoming smaller than one—indicating habitat failure—when the detection time exceeds 470 seconds and the fire spread rate is 1.04 mm/s, defining a critical operational boundary beyond which the habitat’s systems cannot respond effectively. This threshold enables habitat designers using HabSim to set measurable performance requirements for detection and mitigation, directly informing sensor placement, early warning strategies, and resilience-driven design decisions.

Through these simulation results, we demonstrate the utility of the HabSim disruption architecture, which effectively integrates low-computational-cost phenomenological models with more detailed physics-based models. This hybrid modeling approach allows for rapid yet insightful analysis, enabling mission planners and habitat designers to identify critical vulnerabilities and design bottlenecks early in the development process. By striking a balance between computational efficiency and model fidelity, HabSim facilitates iterative testing and optimization, supporting research into resilience and autonomous decision-making across a range of design scenarios. Moreover, the insights from this simulations offer valuable guidance for designing more effective fire detection and suppression strategies. These might include the use of automatic fire suppression systems capable of detecting and extinguishing fires with minimal delay, thereby maximizing response speed and limiting fire propagation. The most effective systems activate within seconds of fire initiation, demonstrating extremely rapid suppression relative to hazard escalation — a performance reflected in a response margin close to 1. In deep space habitats, where crew resources are limited and fires can quickly compromise critical systems, early detection and immediate suppression are essential to preventing cascade failures. Ultimately, this approach provides a powerful tool for advancing the development of resilient, efficient space habitats, particularly under the demanding conditions of long-duration deep space missions.

\section{Conclusions }

This research presents an architecture that integrates physics-based and phenomenological models to simulate fault propagation, detection, and repair in deep space habitats. Implemented in the HabSim testbed, this architecture enables efficient modeling of disruptions by balancing computational efficiency with physical accuracy while supporting real-time execution. 
To facilitate the exploration of resilient design and autonomous operation, the HabSim architecture allows for the integration and propagation of a wide range of faults, followed by repair and recovery processes. The platform supports multiple modeling approaches within a unified architecture, combining low-cost phenomenological models with more detailed physics-based models. While phenomenological models are faster to develop and computationally efficient, physics-based models provide greater accuracy at a higher computational cost. By blending these techniques, our approach ensures sufficient complexity to analyze disruption management and recovery in complex systems. Through an illustrative disruption scenario, we demonstrate and analyze the initiation, propagation, detection, and repair of a fire event using various signal and model types.

This paper also demonstrates the application of the HabSim testbed in evaluating the control effectiveness of a fire suppression system within a space habitat, emphasizing its broader relevance in advancing resilience-based design. By systematically analyzing fire spread rates, detection times, and their effects on recovery time, temperature, and power consumption, we identified key factors influencing a habitat’s resilience to fire hazards. These simulations revealed critical thresholds, beyond which fire propagation becomes uncontrollable, and highlighted the importance of early detection and quick response to avoid irreversible damage. By using the time response margin as a key performance indicator, we support researchers to quantify whether the suppression system  can intervene within the critical time window. 
This approach is essential for developing adaptive methods that optimize response actions, enhance overall habitat resilience to fire disruptions, and identify the most vulnerable aspects of the habitat design, offering greater versatility than traditional, compartmentalized modeling tools. The use of time response margin as a performance metric allowed for the identification of critical thresholds beyond which fire propagation becomes uncontrollable, highlighting the need for efficient and reliable safety controls. These findings not only inform the development of optimized passive and active fire suppression strategies but also underscore the value of HabSim in supporting a wide range of applications, including damage repair prioritization, power system design, environmental control and life support system (ECLSS) design, operational vulnerability assessments, and contingency planning. The architecture's capacity to model disruption propagation offers valuable guidance for scientists, design engineers, and policymakers involved in space habitat development. Looking ahead, future work will further leverage HabSim to enhance autonomous decision-making and address fundamental challenges in long-duration, deep space missions.

\section*{Code Availability}
HabSim, the user manual, and the files to run this sample simulation will be posted on GitHub RETHi repository. The files and data used to generate these results will also be posted. This paper corresponds to a description of version 6.3 of the HabSim simulation code.

\section*{Acknowledgments}
This work was supported by a Space Technology Research Institutes Grant (number 80NSSC19K1076) from NASA’s Space Technology Research Grants Program. The authors would like to thank past and current HabSim subsystem modelers for making this research possible, including Leila Chebbo, Murali Krishnan, Adnan Shahriar, and Seungho Rhee for their direct support.

\setcounter{figure}{0} 
\setcounter{table}{0}  

\begin{landscape}
\appendix
\section{}
\begin{figure}[H]
\centering
\includegraphics[width=1.2\textwidth]{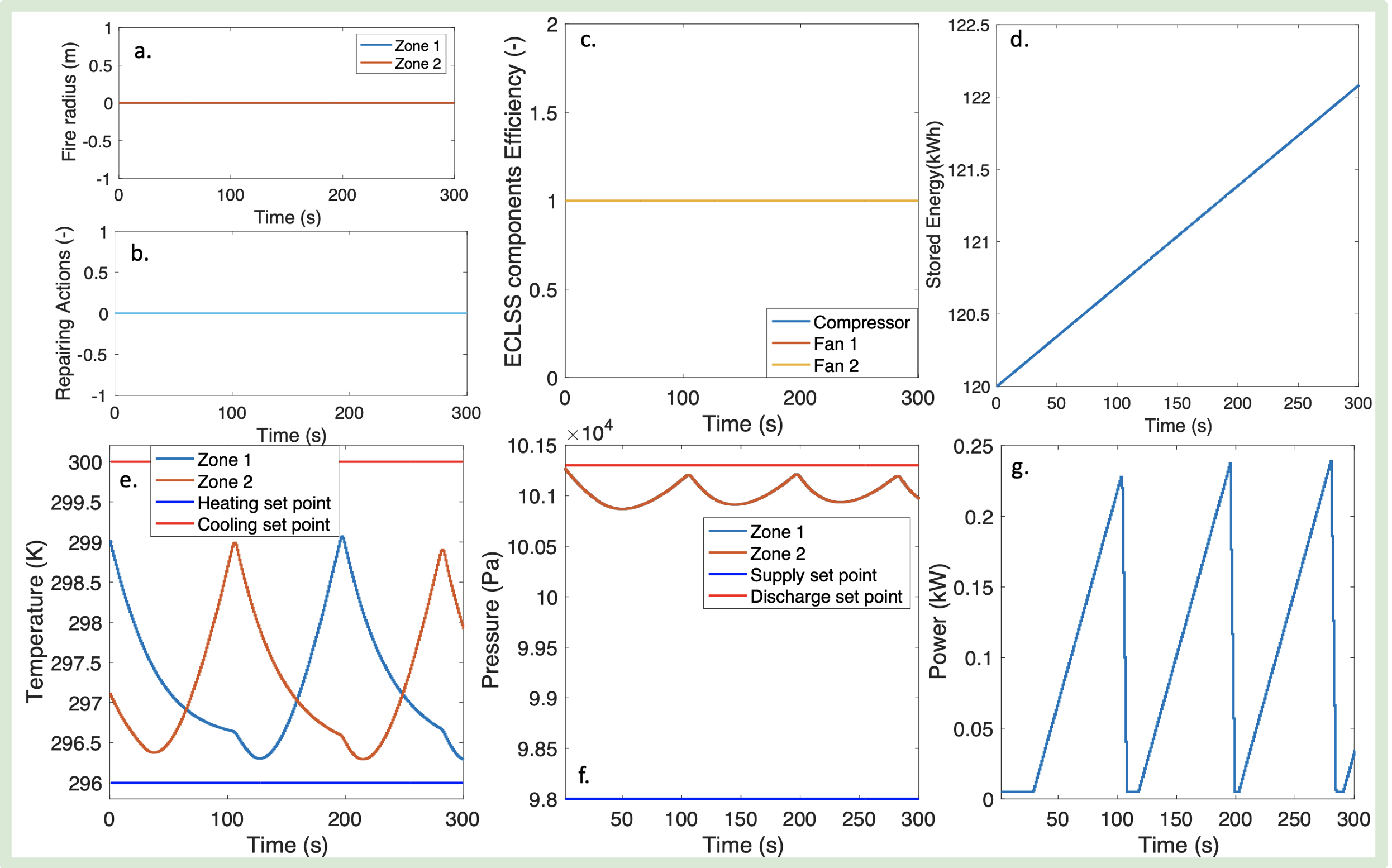}
\caption{Fire scenario data, detail step 1. The habitat is in nominal condition. Pressure and temperature are within the safe range (e, f). The power consumption is due to the thermal and pressure control (g). Fan 1, Fan 2 and compressor have maximum possible efficiency (c). Stored energy is increasing since the batteries are being charged by nuclear and solar power (d). }
\label{fig:Step1}
\end{figure}

\end{landscape}

\begin{landscape}

\begin{figure}[H]
\centering
\includegraphics[width=1.2\textwidth]{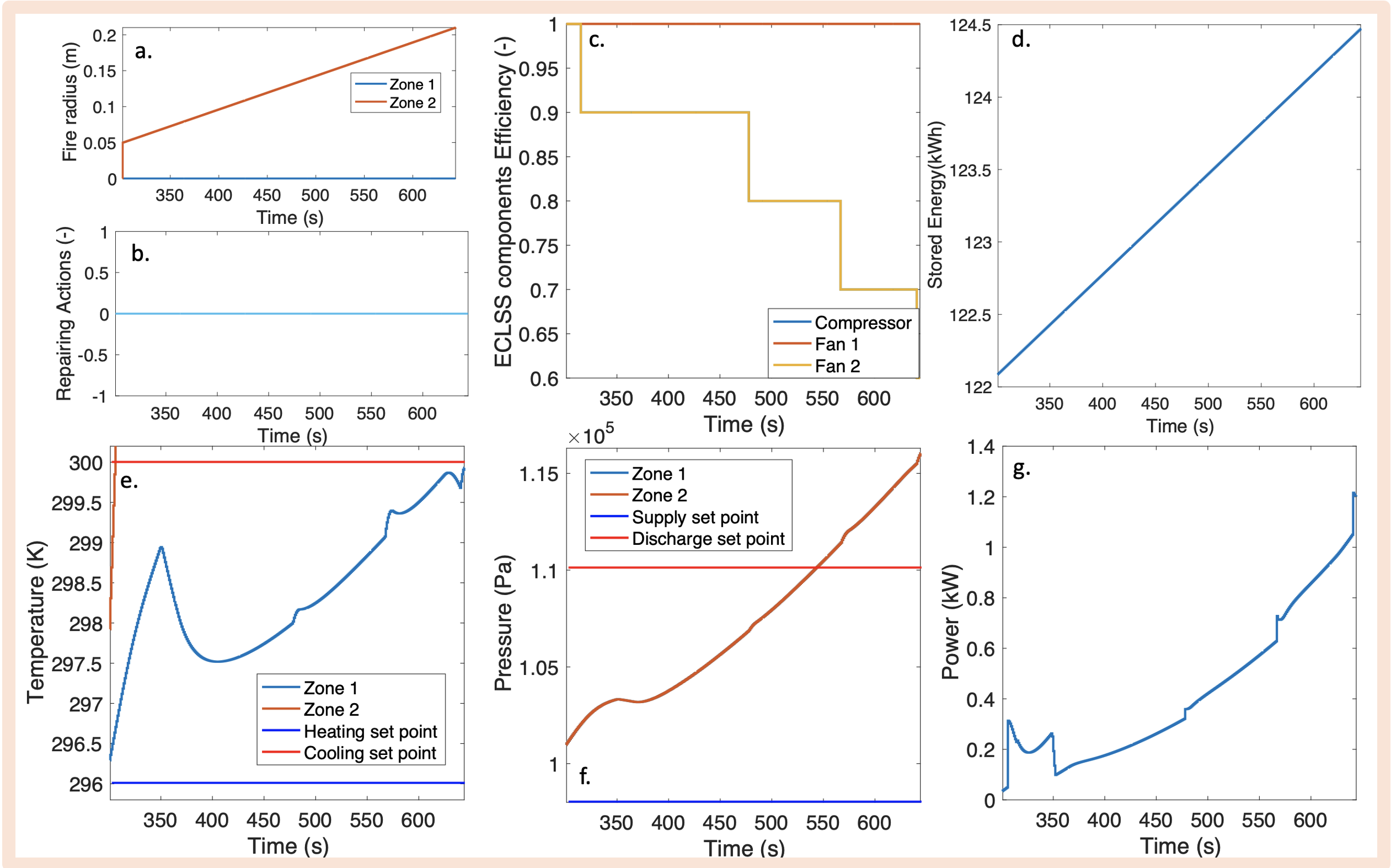}
\caption{Fire scenario data, detail step 2. The fire starts propagating within the habitat (a). Pressure and temperature are increasing, and the pressure exceeds the cooling set point (e, f). The power consumption is due to the thermal and pressure control and starts to increase (g), since Fan 1, Fan 2, and compressor efficiency is decreasing (c). Stored energy is increasing since the batteries are being charged by nuclear and solar power (d), and the power consumption is smaller than the generated power.}
\label{fig:Step2}
\end{figure}

\begin{figure}[H]
\centering
\includegraphics[width=1.2\textwidth]{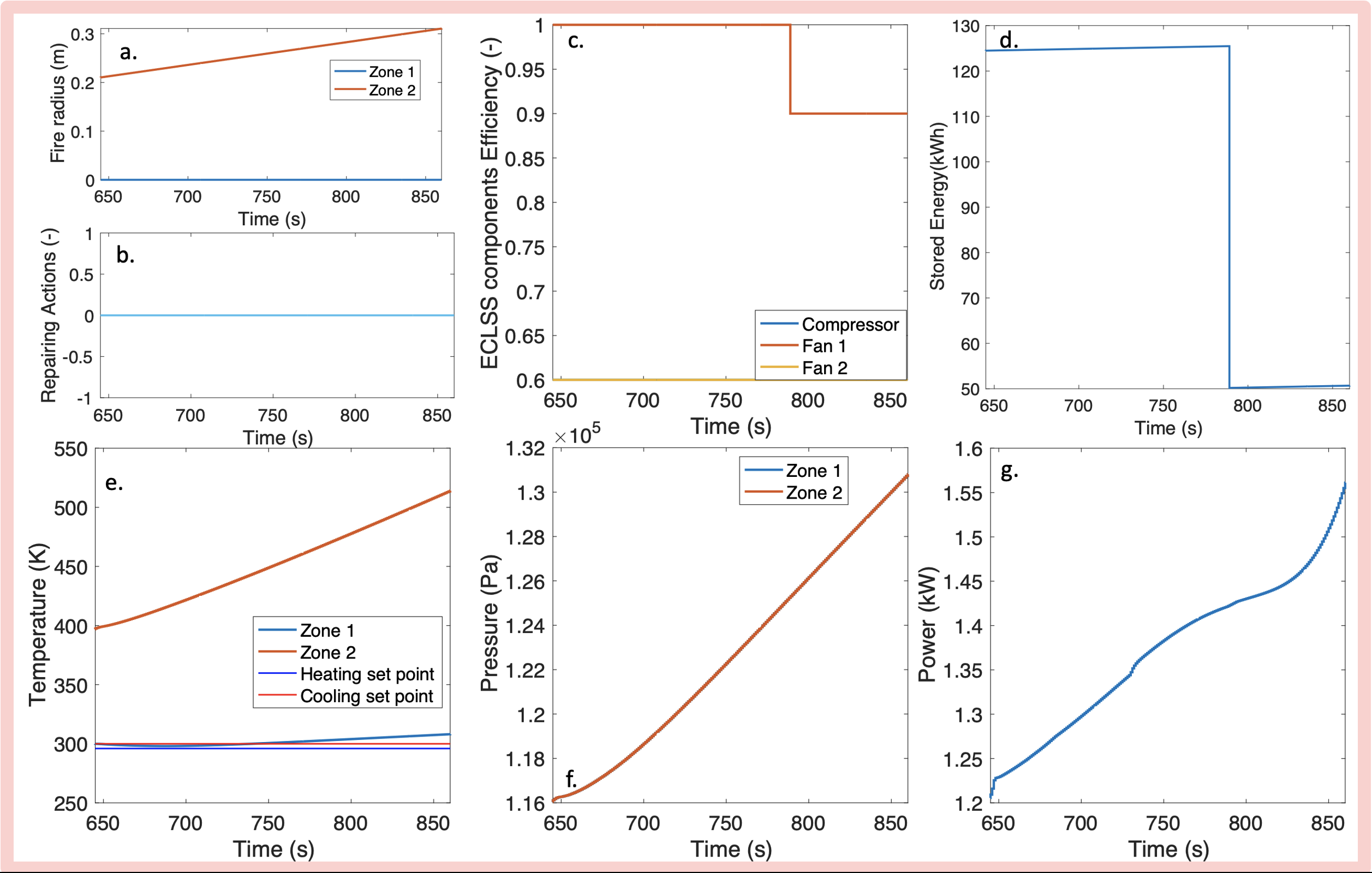}
\caption{Fire scenario data, detail step 3. Pressure in both zones and temperature in zone 2 is outside the safe range (e, f). The power consumption is due to the thermal and pressure control and starts to increase (g), since Fan 1, Fan 2, and compressor efficiency is decreasing (c). Stored energy decreases because the temperature affects its efficiency (d).}
\label{fig:Step3}
\end{figure}

\begin{figure}[H]
\centering
\includegraphics[width=1.2\textwidth]{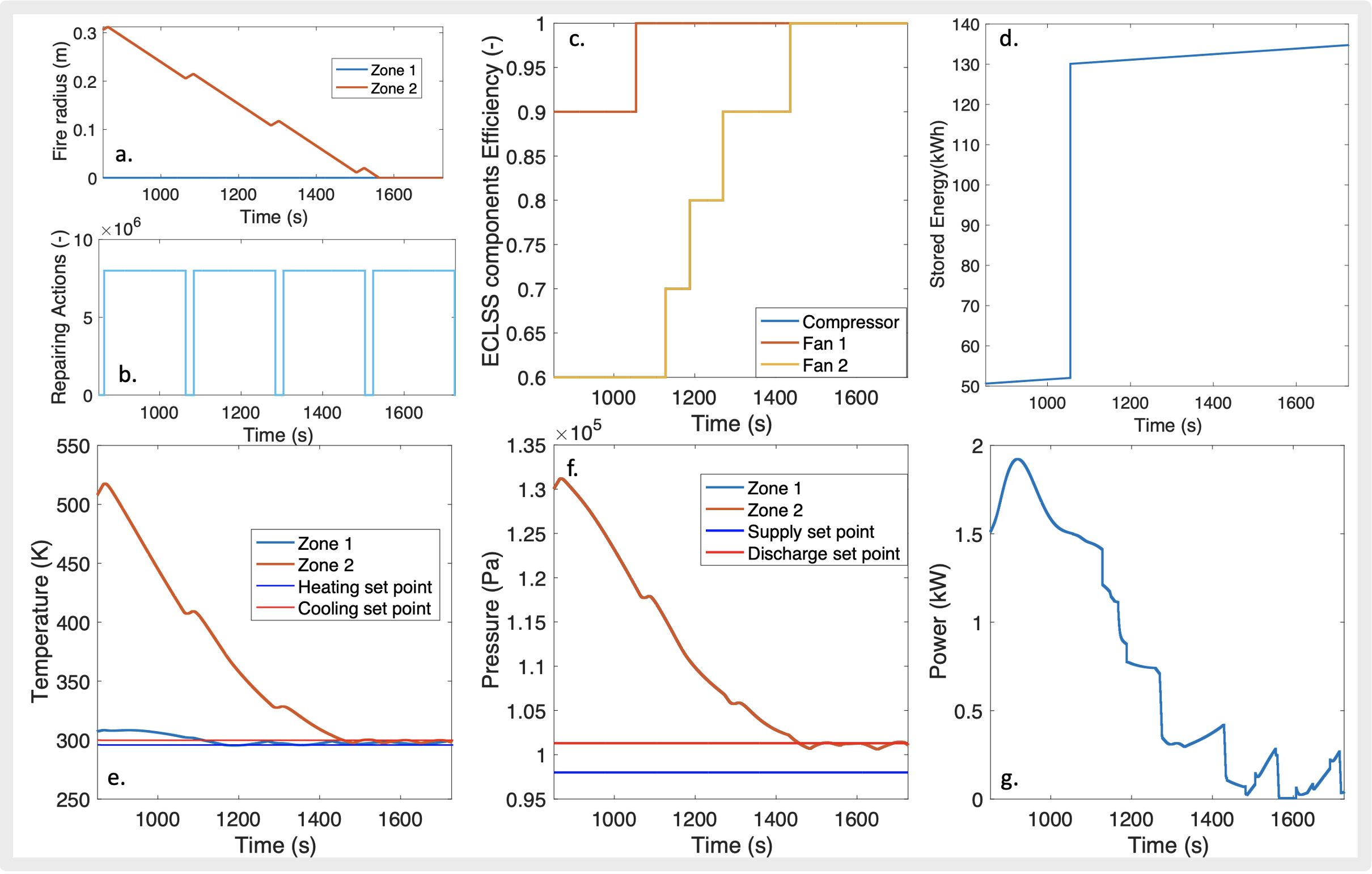}
\caption{Fire scenario data, detail step 4. Pressure in both zones and temperature in zone 2 are still outside the safe range (e, f). The power consumption is due to the thermal and pressure control and starts to decrease (g), since Fan 1, Fan 2, and compressor efficiency is increasing (c). Stored energy increases because the temperature affects its efficiency (d).}
\label{fig:Step4}
\end{figure}

\begin{figure}[H]
\centering
\includegraphics[width=1.2\textwidth]{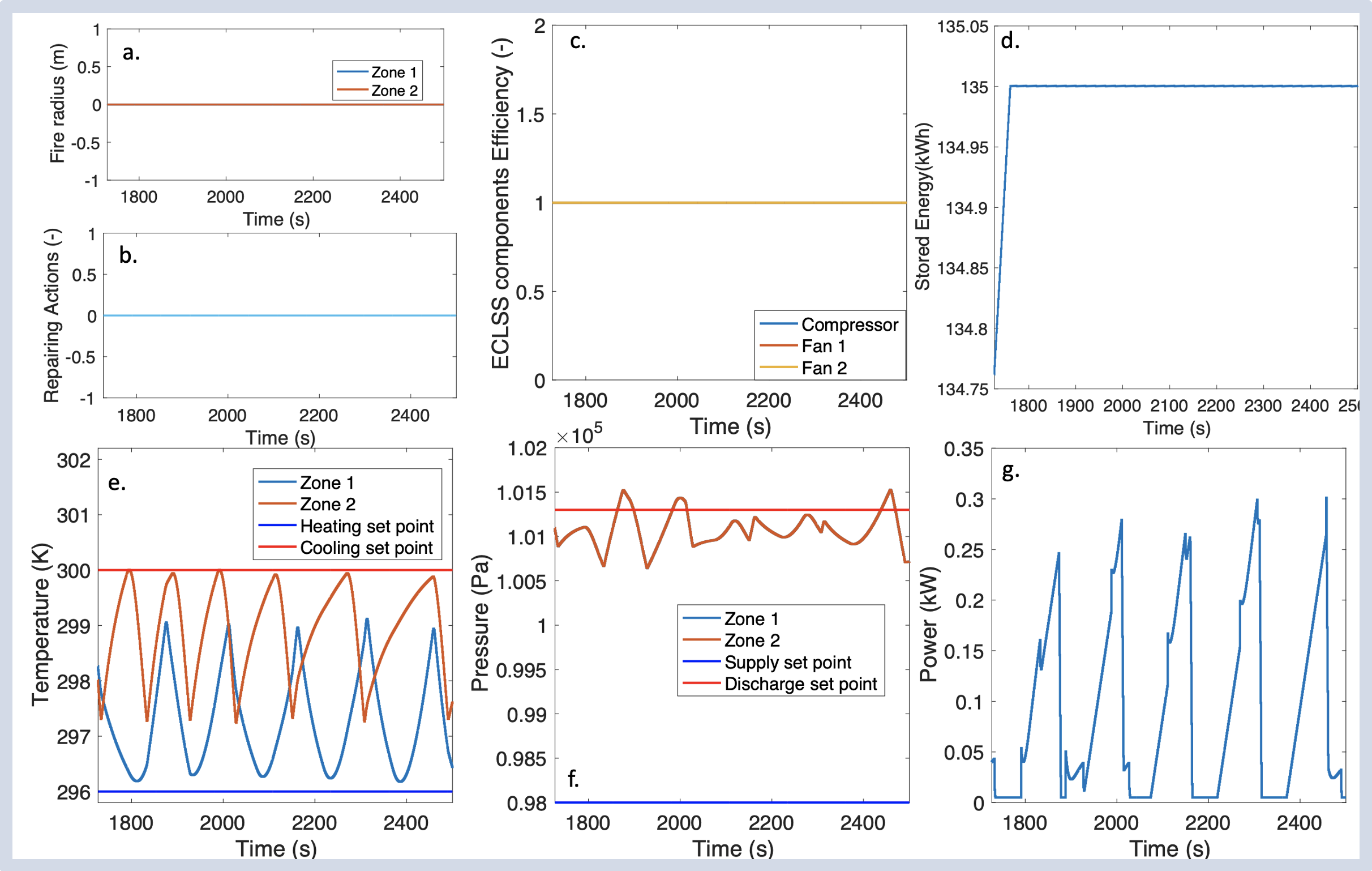}
\caption{Fire scenario data, detail step 5. The habitat is back to nominal conditions. Pressure and temperature are within the safe range (e, f). The power consumption is due to the thermal and pressure control only (g). Fan 1, Fan 2, and compressor are back to maximum possible efficiency (c). Stored energy is increasing since the batteries are being charged by nuclear and solar power (d).}
\end{figure}
\end{landscape}


\bibliographystyle{apalike} 

\bibliography{refs}

\end{document}